\documentclass[aps,pra,twocolumn,showpacs]{revtex4}

\usepackage[dvips]{graphicx}

\input{epsf}

\begin{document}

\title{Equation of state of a Fermi gas in the BEC-BCS crossover: a quantum Monte Carlo study}

\author{G. E. Astrakharchik$^{(1,2)}$, J. Boronat$^{(3)}$, J. Casulleras$^{(3)}$, and S. Giorgini$^{(1)}$}
\address{$^{(1)}$Dipartimento di Fisica, Universit\`a di Trento and BEC-INFM, I-38050 Povo, Italy\\
$^{(2)}$Institute of Spectroscopy, 142190 Troitsk, Moscow region, Russia\\
$^{(3)}$Departament de F\'{\i}sica i Enginyeria Nuclear, Campus Nord B4-B5, Universitat Polit\`ecnica de Catalunya,
E-08034 Barcelona, Spain}

\date{\today}

\begin{abstract}
We calculate the equation of state of a two-component Fermi gas with attractive short-range interspecies interactions using 
the fixed-node diffusion Monte Carlo method. The interaction strength is varied over a wide range by tuning the value $a$ of 
the $s$-wave scattering length of the two-body potential. For $a>0$ and $a$ smaller than the inverse Fermi wavevector 
our results show a molecular regime with repulsive interactions well described by the dimer-dimer scattering 
length $a_m=0.6 a$. The pair correlation functions of parallel and opposite spins are also discussed as a function of 
the interaction strength.            
\end{abstract}


\maketitle

Recent experiments on two-component ultracold atomic Fermi gases near a Feshbach resonance have opened the possibility 
of investigating the crossover from a Bose-Einstein condensate (BEC) to a Bardeen-Cooper-Schrieffer (BCS) superfluid. In 
these systems the strength of the interaction can be varied over a very wide range by magnetically tuning the two-body 
scattering amplitude. For positive values of the $s$-wave scattering length $a$, atoms with different spins are observed to 
pair into bound molecules which, at low enough temperature, form a Bose condensate~\cite{EXP1}. The molecular BEC state is 
adiabatically converted into an ultracold Fermi gas with $a<0$ and $k_F|a|\ll 1$~\cite{EXP2}, where standard BCS theory is 
expected to apply. In the crossover region the value of $|a|$ can be orders of magnitude larger than the inverse Fermi wave 
vector $k_F^{-1}$ and one enters a new strongly-correlated regime known as unitary limit~\cite{EXP3,EXP2}. In dilute 
systems, for which the effective range of the interaction $R_0$ is much smaller than the mean interparticle distance, 
$k_FR_0\ll 1$, the unitary regime is believed to be universal~\cite{Heiselberg,Unitarity}. In this regime, the only 
relevant energy scale should be given by the energy of the noninteracting Fermi gas,
\begin{equation}
\epsilon_{FG}=\frac{3}{10}\frac{\hbar^2k_F^2}{m} \;.
\label{fermienergy}
\end{equation}
  
The unitary regime presents a challenge for many-body theoretical approaches because there is not any obvious small parameter 
to construct a well-posed theory. The first theoretical studies of the BEC-BCS crossover at zero temperature are based on 
the mean-field BCS equations~\cite{BCS-BEC}. More sophisticated approaches take into account the effects of 
fluctuations~\cite{Strinati}, or include explicitly the bosonic molecular field~\cite{Bose-Fermi1}. These theories provide 
a correct description in the deep BCS regime, but are only qualitatively correct in the unitary limit and in the BEC region. 
In particular, in the BEC regime the dimer-dimer scattering length has been calculated exactly from the solution of the 
four-body problem, yielding $a_m=0.6 a$~\cite{Petrov}. Available results for the equation of state in this regime do not 
describe correctly the repulsive molecule-molecule interactions~\cite{Bose-Fermi2}.

Quantum Monte Carlo techniques are the best suited tools for treating strongly-correlated systems. These methods
have already been applied to ultracold degenerate Fermi gases in a recent work by Carlson {\it et al.}~\cite{QMC1}. 
In this study the energy per particle of a dilute Fermi gas in the unitary limit is calculated with the fixed-node 
Green's function Monte Carlo method (FN-GFMC) giving the result $E/N=\xi\epsilon_{FG}$ with $\xi=0.44(1)$. In a 
subsequent work~\cite{QMC2}, the same authors have extended the FN-GFMC 
calculations to investigate the equation of state in the BCS and BEC regimes. Their results in the BEC limit are compatible 
with a repulsive molecular gas, but the equation of state has not been extracted with enough precision.

In the present Letter, we report results for the equation of state of a Fermi gas in the BEC-BCS crossover region using the 
fixed-node diffusion Monte Carlo method (FN-DMC). The interaction strength is
varied over a very broad range from $-6\le -1/k_Fa\le 6$, including the unitary limit and the deep BEC and BCS regimes. 
In the unitary and in the BCS limit we find agreement, respectively, with the results of Ref.~\cite{QMC1} and with the known
perturbation expansion of a weakly attractive Fermi gas~\cite{Huang-Yang}. In the BEC regime, we find a gas of molecules whose 
repulsive interactions are well described by the dimer-dimer scattering length $a_m=0.6 a$. Results for the pair correlation 
functions of parallel and antiparallel spins are reported in the various regimes. In the BEC regime we find agreement with 
the pair correlation function of composite bosons calculated using the Bogoliubov approximation.

The homogeneous two-component Fermi gas is described by the Hamiltonian
\begin{equation}
H=-\frac{\hbar^2}{2m}\left( \sum_{i=1}^{N_\uparrow}\nabla^2_i + \sum_{i^\prime=1}^{N_\downarrow}\nabla^2_{i^\prime}\right)
+\sum_{i,i^\prime}V(r_{ii^\prime}) \;,
\label{hamiltonian}
\end{equation}   
where $m$ denotes the mass of the particles, $i,j,...$ and $i^\prime,j^\prime,...$ label, respectively, spin-up and spin-down
particles and $N_\uparrow=N_\downarrow=N/2$, $N$ being the total number of atoms. We model the interspecies interatomic 
interactions using an attractive square-well potential: $V(r)=-V_0$ for $r<R_0$, and $V(r)=0$ otherwise. 
In order to ensure that the mean interparticle distance 
is much larger than the range of the potential we use $nR_0^3=10^{-6}$, where $n=k_F^3/(3\pi^2)$ is the gas number density. By 
varying the depth $V_0$ of the potential one can change the value of the $s$-wave scattering length, which for this potential 
is  given by 
$a=R_0[1-\tan(K_0R_0)/(K_0R_0)]$, where $K_0^2=mV_0/\hbar^2$. We vary $K_0$ is the range: $0<K_0<\pi/R_0$. For 
$K_0R_0<\pi/2$ the potential does not support a two-body bound state and $a<0$. For $K_0R_0>\pi/2$, instead, the scattering length 
is positive, $a>0$, and a molecular state appears whose binding energy $\epsilon_b$ is determined by the trascendental equation 
$\sqrt{|\epsilon_b|m/\hbar^2}R_0\tan(\bar{K}R_0)/(\bar{K}R_0)=-1$, where $\bar{K}^2=K_0^2-|\epsilon_b|m/\hbar^2$. The value
$K_0=\pi/(2R_0)$ corresponds to the unitary limit where $|a|=\infty$ and $\epsilon_b=0$.    

In a FN-DMC simulation the function $f({\bf R},\tau)=\psi_T({\bf R})\Psi({\bf R},\tau)$, where $\Psi({\bf R},\tau)$ denotes the 
wave function of the system and $\psi_T({\bf R})$ is a trial function used for importance sampling, is evolved in imaginary time 
$\tau=it/\hbar$ according to the Schr\"odinger equation
\begin{eqnarray}
-\frac{\partial f({\bf R},\tau)}{\partial\tau}= &-& D\nabla_{\bf R}^2 f({\bf R},\tau) + D \nabla_{\bf R}[{\bf F}({\bf R})
f({\bf R},\tau)] \nonumber \\
&+& [E_L({\bf R})-E_{ref}]f({\bf R},\tau) \;.
\label{FNDMC}
\end{eqnarray}
In the above equation ${\bf R}=({\bf r}_1,...,{\bf r}_N)$, $E_L({\bf R})=
\psi_T({\bf R})^{-1}H\psi_T({\bf R})$ denotes the local energy, ${\bf F}({\bf R})=2\psi_T({\bf R})^{-1}\nabla_{\bf R}
\psi_T({\bf R})$ is the quantum drift force, $D=\hbar^2/(2m)$ plays the role of an effective diffusion constant, and $E_{ref}$
is a reference energy introduced to stabilize the numerics. The energy and other observables of the state of the system are 
calculated from 
averages over the asymtpotic distribution function $f({\bf R},\tau\to\infty)$. To ensure positive definiteness of the 
probability distribution $f$ for fermions, the nodal structure of $\psi_T$ is imposed as a constraint during the calculation. 
It can be 
proved that, due to this nodal constraint, the calculated energy is an upper bound to the eigenenergy for a given 
symmetry~\cite{FNDMC}. In particular, if the nodal surface of $\psi_T$ were exact, the fixed-node energy would also be exact.  

In the present study we make use of the following trial wave functions. A BCS wave function
\begin{equation}
\psi_{BCS}({\bf R})={\cal A} \left( \phi(r_{11^\prime})\phi(r_{22^\prime})...\phi(r_{N_\uparrow N_\downarrow})\right) \;,
\label{psiBCS}
\end{equation}
and a Jastrow-Slater (JS) wave function
\begin{equation}
\psi_{JS}({\bf R})=\prod_{i,i^\prime}\varphi(r_{ii^\prime}) \left[ {\cal A}\prod_{i,\alpha} 
e^{i{\bf k}_\alpha\cdot{\bf r}_i} \right] \left[ {\cal A}\prod_{i^\prime,\alpha} 
e^{i{\bf k}_\alpha\cdot{\bf r}_{i^\prime}} \right] \;,
\label{psiJS}
\end{equation}
where ${\cal A}$ is the antisymmetrizer operator ensuring the correct antisymmetric properties under particle exchange.
In the JS wave function, Eq. (\ref{psiJS}), the plane wave orbitals have wave vectors ${\bf k}_\alpha=2\pi/L(\ell_{\alpha x}
\hat{x}+\ell_{\alpha y}\hat{y}+\ell_{\alpha z}\hat{z})$, where $L$ is the size of the periodic cubic box fixed by $nL^3=N$, 
and $\ell$ are integer numbers. The correlation functions $\phi(r)$ and $\varphi(r)$ in Eqs. (\ref{psiBCS})-(\ref{psiJS}) 
are constructed from solutions of the two-body Schr\"odinger equation with the square-well potential $V(r)$. In particular, 
in the region $a>0$ we take for the function $\phi(r)$ the bound-state solution $\phi_{bs}(r)$ with energy $\epsilon_b$ 
and in the region $a<0$ the unbound-state solution corresponding to zero scattering energy: 
$\phi_{us}(r)=(R_0-a)\sin(K_0r)/[r\sin(K_0R_0)]$ for $r<R_0$ and $\phi_{us}(r)=1-a/r$ for $r>R_0$. In the 
unitary limit, $|a|\to\infty$, $\phi_{bs}(r)=\phi_{us}(r)$.

\begin{figure}
\begin{center}
\includegraphics*[width=7cm]{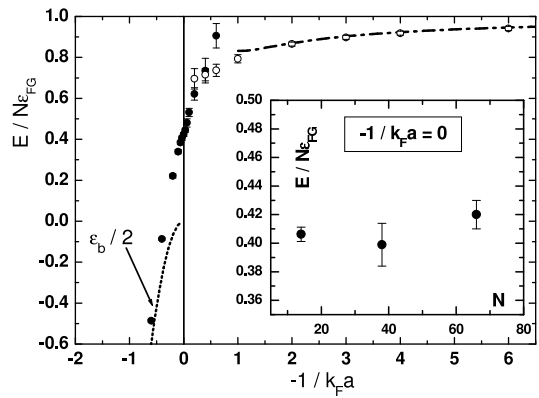}
\caption{Energy per particle in the BEC-BCS crossover. Solid symbols refer to results obtained with the trial wave function 
$\psi_{BCS}$, open symbols refer to the ones obtained with $\psi_{JS}$. The dot-dashed line is the expansion (\ref{BCSexp}) 
holding in the BCS region and the dotted line corresponds to the binding energy $\epsilon_b/2$. Inset: finite size effects in 
the unitary limit $-1/k_Fa=0$.}
\label{fig1}
\end{center}
\end{figure}

The JS wave function $\psi_{JS}$, Eq. (\ref{psiJS}), is used only in the region of negative scattering length, $a<0$, with a 
Jastrow factor $\varphi(r)=\phi_{us}(r)$ for $r<\bar{R}$. In order to reduce possible size effects due to the long range 
tail of $\phi_{us}(r)$, we have used $\varphi(r)=C_1+C_2\exp(-\alpha r)$ for $r>\bar{R}$, with $\bar{R}<L/2$ a matching 
point. The coefficients $C_1$ and $C_2$ are fixed by the continuity condition for $\varphi(r)$ and its first derivative at 
$r=\bar{R}$, whereas the parameter $\alpha>0$ is chosen in such a way that $\varphi(r)$ goes rapidly to a constant. Residual 
size effects have been 
finally determined carrying out calculations with an increasing number of particles $N=14$, 38, and 66. In the inset of 
Fig.~\ref{fig1} we show the dependence of the energy per particle $E/N$ on $N$ in the unitary limit. Similar studies carried 
out in the BEC and BCS regime show that the value $N=66$ is optimal since finite-size corrections in the energy are below 
the reported statistical error in the whole BEC-BCS crossover. We have also checked that effects due to the finite range $R_0$ 
of the potential are negligible.

\begin{figure}
\begin{center}
\includegraphics*[width=7cm]{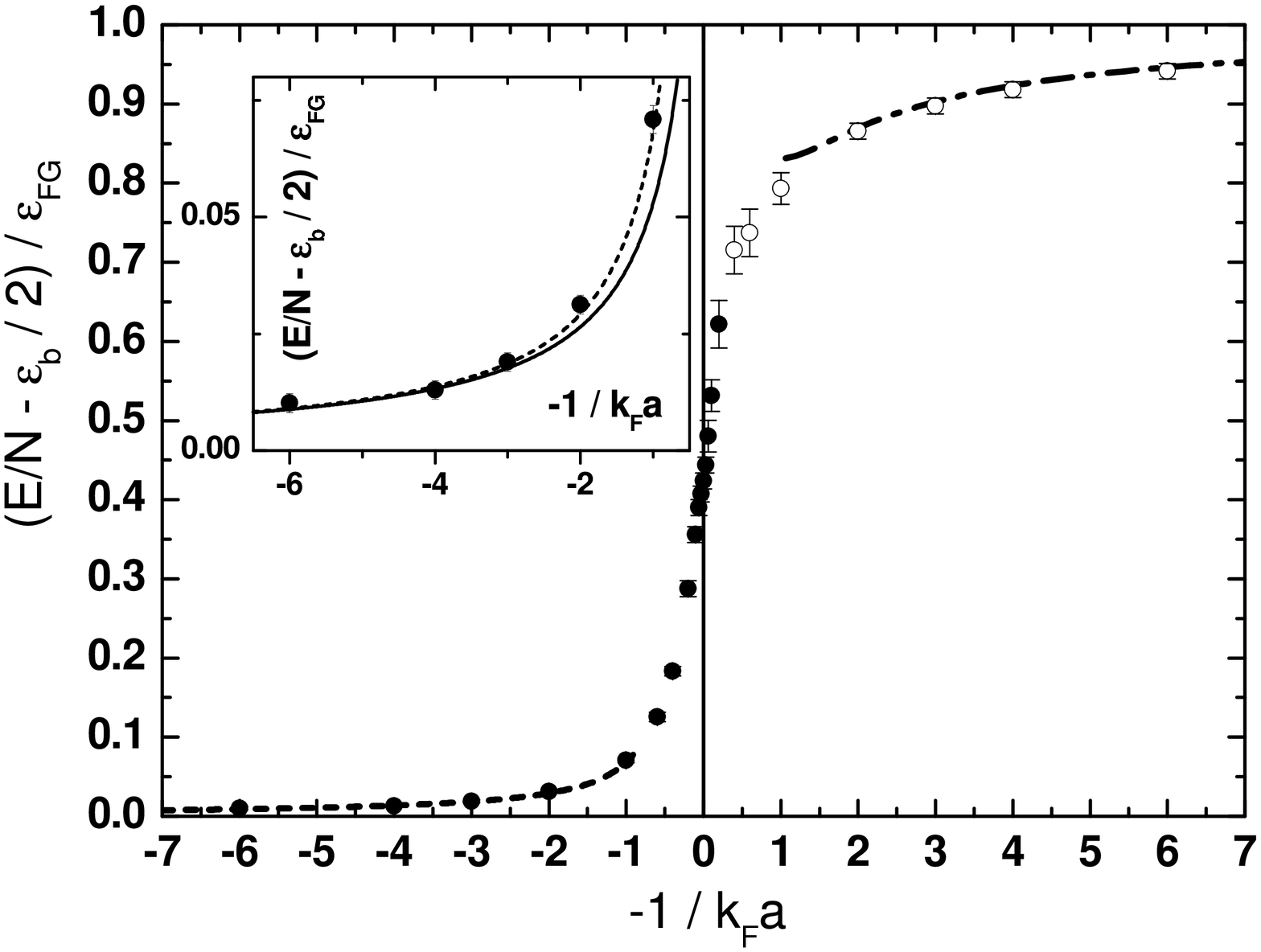}
\caption{Energy per particle in the BEC-BCS crossover with the binding energy subtracted from $E/N$. Solid symbols: results with 
$\psi_{BCS}$, open symbols: results with $\psi_{JS}$. The dot-dashed line is as in Fig.~\ref{fig1} and the dashed line 
corresponds to the expansion (\ref{BECexp}) holding in the BEC regime. Inset: enlarged view of the BEC regime $-1/k_Fa\le-1$. 
The solid line corresponds to the mean-field energy [first term in the expansion (\ref{BECexp})], the dashed line includes the 
beyond mean-field correction [Eq. (\ref{BECexp})].}
\label{fig2}
\end{center}
\end{figure}

\begin{table}
\caption{Energy per particle and binding energy in the BEC-BCS crossover (energies are in units of $\epsilon_{FG}$).}
\begin{tabular}{cccc}
$-1/k_Fa$ & $E/N$ & $\epsilon_b/2$ & $E/N-\epsilon_b/2$ \\ 
\tableline
-6   & -73.170(2)  & -73.1804 & 0.010(2)  \\
-4   & -30.336(2)  & -30.3486 & 0.013(2)  \\
-2   & -7.071(2)   &  -7.1018 & 0.031(2)  \\
-1   & -1.649(3)   &  -1.7196 & 0.071(3)  \\
-0.4 & -0.087(6)   &  -0.2700 & 0.183(6)  \\
-0.2 &  0.223(1)   &  -0.0671 & 0.29(1)   \\
 0   &  0.42(1)    &   0      & 0.42(1)   \\
 0.2 &  0.62(3)    &   0      & 0.62(3)   \\
 0.4 &  0.72(3)    &   0      & 0.72(3)   \\
 1   &  0.79(2)    &   0      & 0.79(2)   \\
 2   &  0.87(1)    &   0      & 0.87(1)   \\
 4   &  0.92(1)    &   0      & 0.92(1)   \\
 6   &  0.94(1)    &   0      & 0.94(1)   \\
\label{tab1}
\end{tabular}
\end{table}

\begin{figure}
\begin{center}
\includegraphics*[width=7cm]{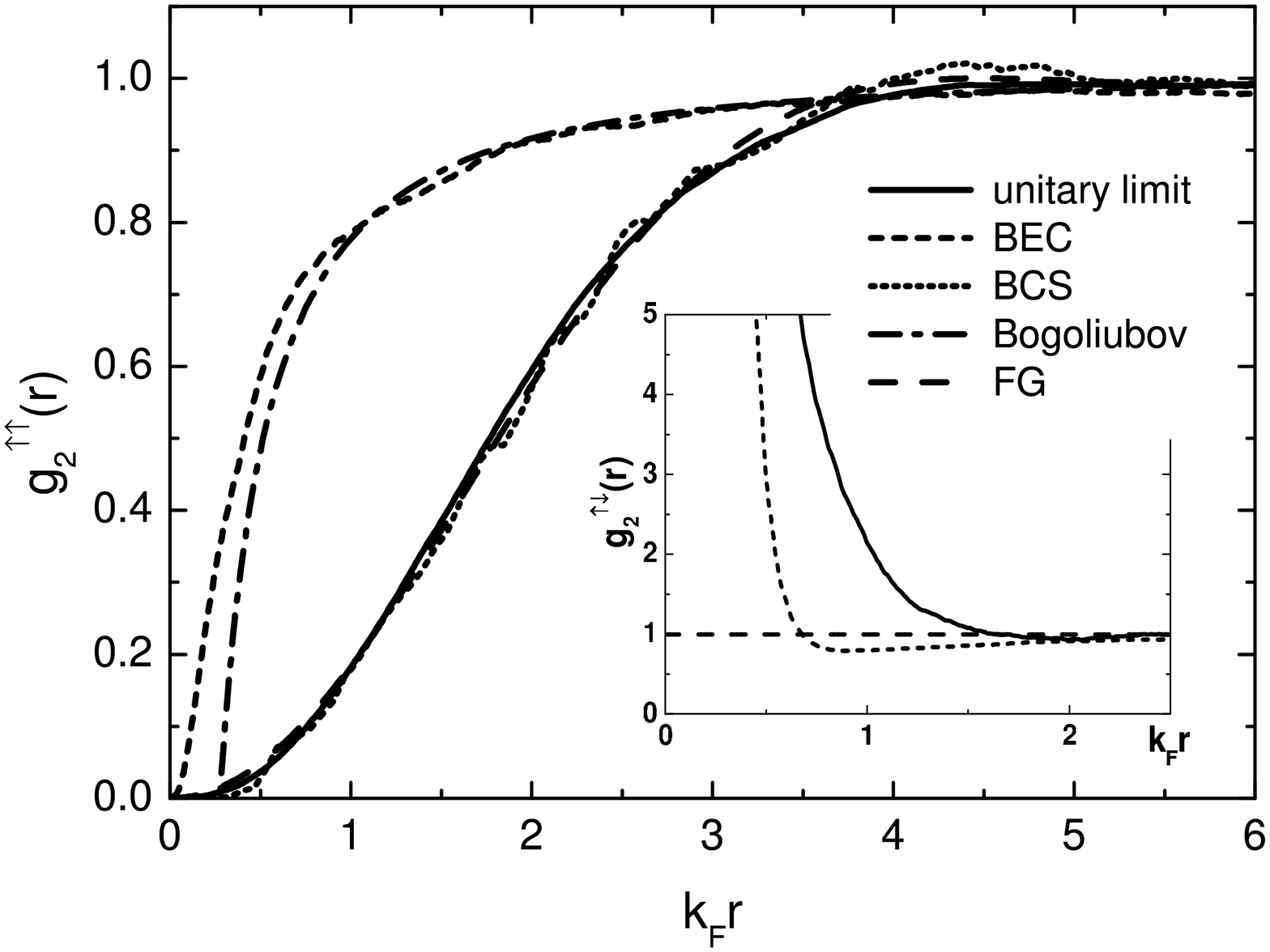}
\caption{Pair correlation function of parallel, $g_2^{\uparrow\uparrow}(r)$, and (inset) of antiparallel spins, 
$g_2^{\uparrow\downarrow}(r)$, for $-1/k_Fa=0$ (unitary limit), $-1/k_Fa=-4$ (BEC regime), $-1/k_Fa=4$ (BCS regime) and for
a noninteracting Fermi gas (FG). The dot-dashed line corresponds to the pair correlation function of a Bose gas with $a_m=0.6a$
and $-1/k_Fa=-4$ calculated using the Bogoliubov approximation.}
\label{fig3}
\end{center}
\end{figure}

The FN-DMC energies for $N=66$ atoms and the potential $V(r)$ with $nR_0^3=10^{-6}$ are shown in Fig.~\ref{fig1} and in 
Table~\ref{tab1} as a function of the interaction parameter $-1/k_Fa$. The numerical simulations are carried out both with the 
BCS wave function, Eq. (\ref{psiBCS}), and with the JS wave function, Eq. (\ref{psiJS}). For 
$-1/k_Fa>0.4$ we find that $\psi_{JS}$ gives lower energies, whereas for smaller values of $-1/k_Fa$, including the unitary 
limit and the BEC region, the function $\psi_{BCS}$ is preferable. This behavior reflects the level of accuracy of the 
variational {\it ansatz} for the nodal structure of the trial wave function. We believe that in the intersection region, 
$-1/k_Fa\sim 0.4$, both wave functions $\psi_{BCS}$ and $\psi_{JS}$ give a poorer description of the exact nodal structure of the 
state, resulting in a less accurate estimate of the energy. In the BCS region, $-1/k_Fa>1$, our results for $E/N$ are in 
agreement with the perturbation expansion of a weakly attractive Fermi gas~\cite{Huang-Yang,note}
\begin{equation}
\frac{E}{N\epsilon_{FG}}= 1+\frac{10}{9\pi}k_Fa+\frac{4(11-2\log2)}{21\pi^2}(k_Fa)^2+... \;.
\label{BCSexp}
\end{equation}              
In the unitary limit we find $E/N=\xi\epsilon_{FG}$, with $\xi=0.42(1)$. This result is compatible with the findings of 
Refs.~\cite{QMC1,QMC2} obtained using a different trial wave function which includes both Jastrow and BCS correlations.
The value of the parameter $\beta=\xi-1$ has been measured in experiments with trapped Fermi gases~\cite{EXP3,EXP2}, but the 
precision is too low to make stringent comparisons with theoretical predictions.  
In the region of positive scattering length $E/N$ decreases by decreasing $k_Fa$. At approximately $-1/k_Fa\simeq-0.3$, the 
energy becomes negative, and by further decreasing $k_Fa$ it rapidly approaches the binding energy per particle $\epsilon_b/2$ 
indicating the formation of bound molecules~\cite{QMC2}. The results with the binding energy subtracted from $E/N$ are shown 
in Fig.~\ref{fig2}. In the BEC region, $-1/k_Fa<-1$, we find that the FN-DMC energies agree with the equation of state of a 
repulsive gas of molecules
\begin{equation}
\frac{E/N-\epsilon_b/2}{\epsilon_{FG}}=\frac{5}{18\pi}k_Fa_m\left[1+\frac{128}{15\sqrt{6\pi^3}}(k_Fa_m)^{3/2}+...\right] \;,
\label{BECexp}
\end{equation}
where the first term corresponds to the mean-field energy of a gas of molecules of mass $2m$ and density $n/2$ interacting 
with the positive molecule-molecule scattering length $a_m$, and the second term corresponds to the first beyond mean-field 
correction~\cite{Lee-Huang-Yang}. If for $a_m$ we use the value calculated by Petrov {\it et al.}~\cite{Petrov} $a_m=0.6a$, 
we obtain the curves shown in Fig.~\ref{fig2}. If, instead, we use $a_m$ as a fitting parameter to our FN-DMC 
results in the region $-1/k_Fa\le-1$, we obtain the value $a_m/a=0.62(1)$. A detailed knowledge of the equation of state of 
the homogeneous system is important for the determination of the frequencies of collective modes in trapped 
systems~\cite{Stringari}, which have been recently measured in the BEC-BCS crossover regime~\cite{EXP4}.   

In Fig.~\ref{fig3} we show the results for the pair correlation function of parallel, $g_2^{\uparrow\uparrow}(r)$, and 
antiparallel spins, $g_2^{\uparrow\downarrow}(r)$. For parallel spins, $g_2^{\uparrow\uparrow}(r)$ must vanish at short 
distances due to the Pauli principle. In the BCS regime the effect of pairing is negligible and $g_2^{\uparrow\uparrow}(r)$ 
coincides with the prediction of a noninteracting Fermi gas $g_2^{\uparrow\uparrow}(r)=1-9/(k_Fr)^4[\sin(k_Fr)/k_Fr-\cos(k_Fr)]^2$. 
This result continues to hold in the unitary limit and it shows that in this regime the Fermi wave vector is not renormalized due 
to interactions, in agreement with the theory of Landau Fermi liquids \cite{Landau}. 
In the BEC regime the static structure factor $S(k)$ of composite 
bosons can be estimated using the Bogoliubov result: $S(k)=\hbar^2k^2/[2M\omega(k)]$, where 
$\omega(k)=(\hbar^4k^4/4M^2+gn_m\hbar^2k^2/M)^{1/2}$ is the Bogoliubov dispersion relation for particles with mass $M=2m$, density 
$n_m=n/2$ and coupling constant $g=4\pi\hbar^2a_m/M$. The pair distribution function $g_2(r)$ of composite bosons, obtained through 
$g_2(r)=1+2/N\sum_{\bf k}[S(k)-1]e^{-i{\bf k}\cdot{\bf r}}$ using the value $a_m=0.6 a$, is shown in Fig.~\ref{fig3} 
for $-1/k_Fa=-4$ and compared with the FN-DMC result. For large distances $r\gg a_m$, where Bogoliubov approximation is expected 
to hold, we find a remarkable agreement. This result is consistent with the equation of state in the BEC regime and shows that 
structural properties of the ground state of composite bosons are described correctly in our approach.     
For antiparallel spins, $g_2^{\uparrow\downarrow}(r)$ exhibits a large peak at short distances due to the 
attractive interaction. In the BEC regime the short range behavior is well described by the exponential decay 
$g_2^{\uparrow\downarrow}(r)\propto\exp(-2r\sqrt{|\epsilon_b|m}/\hbar)/r^2$ fixed by the molecular wavefunction $\phi_{bs}(r)$. 
In the unitary regime 
correlations extend over a considerably larger range compared to the tightly bound BEC regime. In the BCS regime the range of 
$g_2^{\uparrow\downarrow}(r)$ is much larger than $k_F^{-1}$ and is determined by the coherence length $\xi_0=\hbar^2 k_F/(m\Delta)$,
where $\Delta$ is the gap parameter. In this regime the wavefunction we use does not account for pairing and is inadequate
to investigate the behavior of $g_2^{\uparrow\downarrow}(r)$.

In conclusion, we have carried out a detailed study of the equation of state of a Fermi gas in the BEC-BCS crossover using FN-DMC 
techniques. In the BCS regime and in the unitary limit our results are in agreement with known perturbation 
expansions and with previous FN-GFMC calculations~\cite{QMC1,QMC2}, respectively. In the BEC regime, we recover the equation of 
state of a gas 
of composite bosons with repulsive effective interactions which are well described by the molecule-molecule scattering length 
$a_m=0.6a$ recently calculated in Ref.~\cite{Petrov}.      

Acknowledgements: Useful discussions with S. Stringari and L.P. Pitaevskii are gratefully acknowledged. GEA and SG acknowledge 
support by the Ministero dell'Istruzione, dell'Universit\`a e della Ricerca (MIUR). JB and JC acknowledge support from DGI 
(Spain) Grant No. BFM2002-00466 and Generalitat de Catalunya Grant No. 2001SGR-00222.


\begin{thebibliography}{20}
\bibitem{note} Note that for $k_F|a|\ll 1$ the nonanalytic correction to the ground-state energy due to the superfluid gap 
is exponentially small.
\end{thebibliography}
\end{document}